\documentstyle[aps,multicol,prl,psfig]{revtex}

\newcommand{\be}{\begin{eqnarray}}
\newcommand{\ee}{\end{eqnarray}}

\begin{document}

\draft

\title{The diffusion coefficient of propagating fronts with
multiplicative noise}

\author{Andrea Rocco$^{1,2,4}$, Jaume Casademunt$^2$, Ute Ebert$^3$, 
and Wim van Saarloos$^4$}
\address{$^1$Dipartimento di Fisica, Universit\`a di Roma ``La Sapienza'', 
P.le Aldo Moro 2, I-00185 Roma, Italy\\
Istituto Nazionale di Fisica della Materia, Unit\`a di Roma\\
$^2$Departament ECM, Universitat de Barcelona, Av. Diagonal 
647, E-08028, Barcelona, Spain\\
$^3$Centrum voor Wiskunde en Informatica, Postbus 94079, 1090 
GB Amsterdam, The Netherlands\\
$^4$Instituut--Lorentz, Universiteit Leiden, Postbus 9506,
2300 RA Leiden, The Netherlands}

\date{\today} 

\maketitle

\begin{abstract}    
Recent studies have shown that in the presence of noise both fronts
propagating into a metastable state and so-called pushed fronts
propagating into an unstable state, exhibit diffusive wandering about
the average position. In this paper we derive an expression for the
effective diffusion coefficient of such fronts, which was motivated before
on the basis of a multiple scale ansatz. 
Our systematic derivation is based on the decomposition 
of the fluctuating front into a suitably positioned average profile
plus fluctuating eigenmodes of the stability 
operator. While the fluctuations of the front position in this 
particular decomposition are a Wiener process on all time scales, 
the fluctuations about the time averaged front profile relax exponentially. 
\end{abstract} 


\begin{multicols}{2}

\section{Introduction}

One of the aspects of front propagation that have been studied in the
literature in recent years is the effect of fluctuations on
propagating fronts \cite{lemar,breuer,armeroprl,armero}. 
In particular it has been found that in the presence of
noise both one-dimensional fronts between a stable and a metastable
state  (``bistable fronts'') and so-called {\em pushed} fronts 
which propagate into an unstable state \cite{pulled}, 
exhibit a diffusive wandering about their average position
\cite{armero}.
This contrasts with the fluctuation behavior of so-called {\em pulled}
fronts propagating into an unstable state which is subdiffusive 
\cite{rocco}. In this paper we shall consider only the case
of pushed  and bistable fronts, however.

Recently, Armero {\em et al.} \cite{armero} derived an expression for the 
effective
diffusion coefficient of a pushed front in the stochastic field equation  
\be
\frac{\partial \phi}{\partial t} = \frac{\partial^2 \phi}{\partial x^2} 
+ f(\phi)
+ \varepsilon^{1/2} g(\phi)\eta(x,t) \label{maineq}
\ee
with a noise term whose correlations are
\be
\overline{\eta(x,t)} = 0, \label{corr1}
\ee
\be
\overline{\eta(x,t) \eta (x^{\prime},t^{\prime})} = 2
C(|x-x^{\prime}|/\Lambda) \delta(t-t^{\prime}). \label{corr2}
\ee
In (\ref{maineq}), $f$ is a nonlinear function 
 of the
field $\phi$ with a stable state at $\phi = 1$ and either a (meta)stable 
or unstable state at $\phi = 0$ and $g(\phi)$ is some other general
nonlinear function. In (\ref{corr1}) and
(\ref{corr2}) the overbar denotes an average over the realizations of
the noise. In order that our noise of Stratonovich type is
well-defined, we have introduced a spatial cutoff in the noise
correlation function (\ref{corr2}) (see \cite{armero} for further
details). 

The derivation in \cite{armero} of the effective front diffusion
coefficient $D_f$ relied on a small-noise stochastic multiple scale
analysis that was based on the idea that the mean square displacement
of the front about its average position was slow relative to the
deterministic relaxation of the front. The basic idea was that only 
the low-frequency components of the noise are responsible for the 
front wandering, so that the high-frequency components, which renormalize 
the front shape and its velocity, could be 
implicitly integrated out. This led to an ansatz for the relative 
scaling of fast and slow time variables where the small parameter 
governing the separation of time scales was the diffusion coefficient 
$D_f$ of the front itself. 
The method then selfconsistently provided an explicit prediction for $D_f$.
The main weakness of the approach was that the above coarse-graining  
procedure could not be carried out explicitly, 
since while there is a separation of time scales for the {\em average} 
quantities, a scale separation scheme is not natural 
for the {\em fluctuating} quantities. 
Hence the derivation had to rely on an uncontrolled ansatz.
In this Brief Report we therefore reconsider this problem. 
We justify the previously derived result for $D_f$ with a 
systematic small-noise expansion based on decomposing the motion of
the 
front into a  diffusive motion of the properly defined front position  plus
fluctuations about the average front profile.
 Technically the fluctuating front position is defined by 
requiring that the fluctuations about the mean front profile are 
orthogonal to the (left) translation mode.
This derivation shows that the previous multiple scale ansatz is not 
quite adequate, and it will 
clarify the connection between the separation of time scales invoked 
in Ref.\cite{armero}, the small noise expansion and the existence of 
a finite gap in the linearized evolution operator. 
The key point of our new derivation is the fact that there is a
unique choice for the collective coordinate $X(t)$ of the 
front profile to be a memoryless Markovian process, and that  
the fluctuations about the average profile then relax exponentially.
This relaxation can be deduced from the spectrum of the linearization 
operator about the average front profile.
In addition our method provides a general strategy to address 
the problem of fluctuations of fronts and other coherent structures.

\section{Derivation of the effective diffusion coefficient}

We can rewrite Eq.\ (\ref{maineq}) in terms of a noise term $R$ 
whose average $\overline{R}$ is zero and a deterministic renormalized part,
\be
\frac{\partial \phi}{\partial t} = \frac{\partial^2 \phi}{\partial x^2} 
+ h(\phi)
+ \varepsilon^{1/2}\;R(\phi,x,t), \label{ito}
\ee
using Novikov's Theorem, as discussed in \cite{armero}. In Eq.\ (\ref{ito}), 
\be
&&h(\phi) = f(\phi) + \varepsilon C(0) g^{\prime}(\phi) g(\phi), \\
&&R(\phi,x,t) = g(\phi) \eta(x,t) - \varepsilon^{1/2} C(0) 
g^{\prime}(\phi) g(\phi),
\ee
where $C(0)$ is of order $\Lambda^{-1}$, so that Eq. (\ref{corr2}) 
yields a delta correlation in space in the limit $\Lambda \rightarrow 0$
\cite{note}. The main idea of the derivation is to introduce a collective
coordinate $X(t)$ for the position of the front. Of course there are
various choices for the position $X(t)$, but as we shall show a
particular choice makes the equations quite transparent. We decompose
the fluctuating field $\phi$ as 
\be
\phi = \phi_0(\xi - X(t)) + \phi_1(\xi - X(t),t). \label{decomp}
\ee
Here $\phi_0$ is the solution of the ODE for the shape of a deterministic
front with velocity $v_R$, the velocity of the deterministic front
associated with Eq.\ (\ref{ito}) with $R=0$ (the subscript $R$ on $v_R$ 
reminds us that the front speed is determined by $h(\phi)$ rather than 
$f(\phi)$, and hence is renormalized due to the presence of the
noise). In other words, $\phi_0$ 
satisfies
\be
0 = \frac{d^2 \phi_0(\xi)}{d \xi^2} + v_R \frac{\partial
\phi_0(\xi)}{\partial \xi} + h(\phi_0). \label{ode}
\ee
While $\phi_0$ is a non-fluctuating quantity, $\phi_1$ is a stochastic
field which contains the fluctuations
about $\phi_0$. In the above, $\xi = x - v_R t$ is the proper variable
for a deterministic front moving with the asymptotic velocity $v_R$,
but note that in (\ref{decomp}) the fields are written in terms of the
shifted variable 
\be
\xi_X = \xi - X(t) = x - v_R t + X(t),
\ee
where $X(t)$ is the rapidly fluctuating  front position whose
explicit definition in terms of a spatially averaged front profile is given below.

As is well known, the derivation of a moving boundary approximation
for deterministic equations (see {\em e.g.} \cite{karma,ebert2} 
and references therein) normally proceeds by projecting onto the
zero mode. Indeed associated with the front solution $\phi_0$ of
(\ref{ode}) is a zero mode of the stability operator
\be
{\cal L} = \frac{\partial^2}{\partial \xi^2} + v_R
\frac{\partial}{\partial \xi} + h^{\prime}(\phi_0),
\ee
which is obtained by linearizing about $\phi_0$. This zero mode
expresses translational invariance, and indeed implies that 
\be
{\cal L} \Phi_R^{(0)} = 0 \;\;\;\;\; \Leftrightarrow  \;\;\;\;\;
\Phi_R^{(0)} = \frac{d\phi_0}{d \xi}. \label{right}
\ee
In our case the operator ${\cal L}$ is not self-adjoint, since $v_R
\neq 0$; as a result the left eigenmode  $\Phi_L^{(0)}$ is different
from $\Phi_R^{(0)}$, but it is known to be (see {\em e.g.} 
\cite{armero,ebert2}):
\be
{\cal L}^{+} \Phi_L^{(0)} = 0  \;\;\;\;\; \Leftrightarrow  \;\;\;\;\; 
\Phi_L^{(0)} = e^{v_R \xi} \frac{d \phi_0}{d \xi}. \label{left}
\ee

As we mentioned above, a particular definition of the position $X(t)$
is especially convenient: we take $X(t)$ defined implicitly by the
requirement that the fluctuating field $\phi_1$ is orthogonal to the
left zero mode. Indeed, defining
\be
\langle A(\xi) B(\xi) \rangle = \int _{-\infty}^{\infty} d \xi A(\xi) B(\xi),
\ee
we require 
\be
&&\langle \Phi_L^{(0)} \phi_1(\xi,t) \rangle \nonumber \\
&& \qquad = \int d \xi e^{v_R \xi} \frac{d \phi_0}{d \xi} \big(\phi 
- \phi_0(\xi - X(t)\big) = 0.  
\ee
Note that at any moment, the {\em fluctuating} front position $X(t)$
is defined in terms of {\em weighted spatial average} of the
fluctuating field $\phi$. 

Upon substitution of Eq.\ (\ref{decomp}) into (\ref{ito}) and
linearization in $\phi_1$ (which is justified for small noise), we
obtain
\be
\frac{\partial \phi_1}{\partial t} = {\cal L} \phi_1 - \dot{X}(t)
\frac{\partial \phi_0}{\partial \xi_X} + R(\phi_0,\xi,t). \label{dtp}
\ee
Note that we have also approximated $R(\phi,\xi_X,t)$ by
$R(\phi_0,\xi,t)$,
which again is correct to lowest order in the noise. 

In addition to the zero mode, the operator ${\cal L}$ will in general
have right eigenmodes $\Phi_R^{(l)}$ with eigenvalues $-\sigma_l$
\be
{\cal L} \Phi_R^{(l)} = -\sigma_l \Phi_R^{(l)}, \hspace{1cm} l \neq 0 
\ee
and with associated left eigenfunctions $\Phi_L^{(l)} = e^{v_R \xi} 
\Phi_R^{(l)}$. Our convention to have the eigenvalues $-\sigma_l$ 
anticipates that
the dynamically relevant front solution is stable, so that all
eigenvalues $\sigma_l$ are positive. Moreover both for fronts between
a stable and a metastable state and for pushed fronts propagating into
an unstable state, the spectrum is known to be gapped \cite{ebert,wim}, 
{\em i.e.} the smallest eigenvalue is strictly greater than zero 
\cite{ebert,wim}.

Since $\phi_1$ is orthogonal to $\Phi^{(0)}_L$, 
we can expand $\phi_1$ in terms of the eigenmodes $\Phi_R^{(l)}$
($l \ge 1$) of ${\cal L}$ as 
\be
\phi_1(\xi_X,t) = \sum_{l \ne 0} a_l(t) \Phi_R^{(l)}(\xi_X). \label{expansion}
\ee 
Substitution of this expansion into (\ref{dtp}) then yields upon
projection onto the zero mode $\Phi_L^{(0)}$ 
\be
\dot{X}(t) = \varepsilon^{1/2} \frac{\langle \Phi_L^{(0)}
R(\phi_0,\xi,t)\rangle} {\langle\Phi_L^{(0)}\Phi_R^{(0)}\rangle}. \label{xdot}
\ee
Taking the square of this result, integrating and averaging 
over the noise,
\be
\overline{X^2(t)} = 2 D_f t = \int_0^t \! dt^{\prime} \int_0^t \! dt^{\prime \prime}
\; \overline{\dot{X}(t^{\prime})\dot{X}(t^{\prime\prime})},
\ee
then yields with (\ref{corr2}), (\ref{right}) and (\ref{left})
\be
D_f = \varepsilon \frac{\int d\xi e^{2 v_R \xi} (d\phi_0 / d \xi)^2 
g^2(\phi_0)}{[\int d\xi e^{v_R \xi} (d\phi_0 / d \xi)^2]^2}. \label{diff}
\ee

This is precisely the result derived earlier in \cite{armero}.
To lowest order in the present small-noise expansion, the average 
front profile is simply $\phi_0$. However, notice that $\phi_0$ 
contains a dependence on $\Lambda$ through $C(0)$ in 
$h(\phi)$. The parameter $C(0)$ must be considered as
an independent one, so that the result (\ref{diff}) has to be interpreted 
as to first order in $\varepsilon$ but to all orders in $\varepsilon / \Lambda$.

The above derivation allows us to also
obtain the relaxation of a fluctuation about the average. Indeed, upon
substituting (\ref{expansion}) into (\ref{dtp}) and projecting onto
the left zero modes, using $\langle \Phi_L^{(n)} \Phi_R^{(m)}\rangle =
\delta_{nm}$ for normalized eigenmodes, we obtain to lowest order
\be
\frac{d a_l}{dt} = - \sigma_l a_l + \varepsilon^{1/2} \langle
\Phi_L^{(l)} R \rangle
\ee
as terms $\dot X(t) d \phi_1/d\xi$ are of higher order in $\varepsilon$.
Note that each mode is damped and has its noise strength weighted 
by $\Phi_L^{(l)}$. One can derive from here in a
straightforward way the mean square fluctuations about the average 
profile. 

We finally note that our discussion clarifies the difficulty of using
a separation of time scales argument for the derivation of the effective diffusion
coefficient: the collective coordinate $X(t)$ is a memoryless Markov
process, and hence the changes in the position have zero correlation
time while  the average of $X^2(t)$ changes slowly. The
coefficients $a_l(t)$,  on the other hand,  have a finite correlation
time and hence are correlated on timescales in between the one of
instantaneous position $X(t)$  and the mean square wandering $\overline{X^2(t)}$.

\section{Concluding Remarks}

We have reported an improved derivation of the diffusion coefficient 
of propagating pushed fronts with multiplicative noise, previously 
found in Ref.\cite{armero}. The present  derivation is more transparent 
and elegant since it is fully explicit and based on standard  
projection techniques. The key point is the identification of a 
definition of the front position which naturally implies the 
diffusive wandering of the front, and avoids invoking an uncontrolled  
hypothesis in addition to the basic assumption of small noise strength. 
This has also clarified that the time scale separation used in 
Ref.\cite{armero} can be traced back to the small noise approximation 
together with the existence of a finite gap in the spectrum of 
the linearized evolution operator.
All these considerations can be
generalized to the effect of fluctuations on other types of 
coherent structures. 

Our derivation of the solvability expression (\ref{diff}) for $D_f$ of
a propagating front shows that the collective coordinate $X(t)$
responds instantaneously to the noise $R$: There is no memory term in
(\ref{xdot}), so that $X(t)$ is Markovian and, more precisely, 
it coincides with the Wiener process (to lowest order in the noise strength).
We stress that this is only true for our particular
definition of $X(t)$ in terms of the orthogonality of $\phi_1$ to the
left zero mode. For any other definition, like the usual one to
define the front position as $X(t) = \int d\xi \; \phi(\xi)$,  
$X(t)$ will not be a Markov process, and would show only 
diffusive behavior at sufficiently long time scales.

As a byproduct of our derivation we have also obtained an explicit expression
for the relaxation behavior  of the fluctuations about the mean front profile.
Not surprisingly, the larger the gap in the spectrum, the
faster the relaxation. As is well known, in models in which there is
a transition from the pushed regime to the pulled regime, the gap
closes upon approaching the transition from the pushed side
\cite{ebert}. Hence the relaxation becomes  slower and slower. As is discussed in \cite{ebert}, in
the pulled regime the spectrum is gapless and this leads to anomalous
power law relaxation of deterministic fronts towards their asymptotic
speed and shape. As a result pulled fronts cannot be described by a moving
boundary approximation \cite{ebert2} and in the presence of
fluctuations they exhibit subdiffusive wandering \cite{rocco} in one
dimension and anomalous scaling in higher dimensions \cite{goutam1,goutam2}.

We are grateful to L. Ram\'\i rez-Piscina for illuminating discussions.
Financial support from TMR network project ERBFMRX-CT96-0085 is acknowledged. 
J.C. also acknowledges financial support from project BXX2000-0638-C02-02.

\end{multicols}


\begin{thebibliography}{99}
\bibitem{pulled}  Deterministic fronts which propagate into a linearly
unstable state are called pulled if their asymptotic speed $v_{as}$
equals the asymptotic spreading speed $v^*$ of linear perturbations
about the unstable state: $v_{as}=v^*$. For pushed fronts, $v_{as}
>v^*$. For fronts propagating into a metastable state, $v^*=0$. See
e.g. \cite{ebert} and references therein for further details.
\bibitem{note} One takes $C(0)$ of the order $\Lambda^{-1}$ 
in order to have $C(x/\Lambda)$ converge to a delta function
and $\int dx C(x/\Lambda) = 1$ \cite{armero}.
\end{thebibliography}
\end{document}